\documentclass[12pt]{iopart}
\usepackage{amssymb}

\begin{document}

\title[ ]{Scalar-tensor theories and asymmetric resonant cavities.}
\author{F. O. Minotti}
\address{Departamento de F\'{\i}sica, Facultad de Ciencias Exactas y Naturales,
Universidad de Buenos Aires - Instituto de F\'{\i}sica del Plasma (CONICET) }

\begin{abstract}
Recently published experimental results indicate the appeareance of unusual
forces on asymmetric, electromagnetic resonant cavities. It is argued here
that a particular class of scalar-tensor theories of gravity could account
for this effect.
\end{abstract}

\maketitle

\section{Introduction}

Very recently, experimental results were published\cite{juan} on the
measurement of forces on closed, asymmetric electromagnetic resonant
cavities. These results add to previous claims in the same line by an
independent researcher who did the first experiments\cite{emdrive}. Such
claims were criticized by the scientific community mainly due to the
proposed theoretical explanation, as Maxwell equations and Special
Relativity clearly indicate that no force is possible without the emission
of radiation from the cavity. On the other hand, it appears that General
Relativity might allow for such kind of reaction-less propulsion, as
exemplified and noted for the first time in\cite{lobo}, where the low
velocity limit of some warp drive spacetimes was analyzed. As indicated
there, negative energy densities are required to accomplish that and,
notably, some scalar fields present this possibility\cite{barcelo}. Of
course, in order to have measurable effects similar to those reported, the
coupling of the scalar field to matter or other fields acting as its source
should be sufficiently strong, and this is precisely what has been proposed
in\cite{MLR} for the case of the electromagnetic field as source of the
scalar field to explain discordant measurements of Newton gravitational
constant. It is then only natural to wonder whether that theory (or a
similar one) may account for the forces reported in resonant cavities. Of
course, all this is highly speculative, and more prosaic explanations for
these forces should be considered first. We proceed on the assumption that
all spurious effects were accounted for, and on the belief that the
possibility presented here is worth exploring.

The theory put forward by Mbelek and Lachi\`{e}ze-Rey in\cite{MLR} (see also 
\cite{Mb}) represents a reduction to four dimensions of a Kaluza-Klein
theory coupled to an external scalar $\psi $, which in turn couples to
matter. It is the source term of $\psi $ which allows for a possible strong
coupling of the Kaluza-Klein scalar $\phi $ to other fields, in particular
to the electromagnetic field. The theory was applied in cosmological\cite%
{mbelek2003} and galactic situations\cite{mbelek2004} and, as mentioned, it
was also used to investigate the possibility of the Earth's magnetic field
influencing the measurements of Newton gravitational constant. In all these
applications of the theory only its weak-field limit was used, and as this
limit is similar for a wide range of theories, we employ in this work a
rather general scalar-tensor theory, which incorporates the additional
external scalar $\psi $.

In the next sections the equations of the mentioned scalar-tensor theory are
derived from its proposed action, along with the equation of motion of
neutral matter. Some axisymmetric electromagnetic modes of a truncated
conical cavity are then presented and used as source in the weak-field
approximation of the equations, previously obtained, to determine the force
on the cavity. It is found that a coupling of the same magnitude as used in 
\cite{MLR} between the scalar $\phi $ and the electromagnetic field results
in a correct magnitude and sign for the forces reported in asymmetric
resonant cavities. As expected, the solution for the cavity presents
negative energy densities (more precisely, it violates the weak energy
condition\cite{hawking}). The theory, however, does not seem to be
completely satisfactory because in its linearized version it also predicts
strong gravitational effects by the Earth's magnetic field, which are
clearly not observed. A possible resolution of this problem is considered in
the last section.

\section{Scalar-tensor theory}

We will consider a scalar-tensor theory of the Brans-Dicke type\cite%
{bransdicke} with inclusion of a Bekenstein's direct interaction of scalar
and Maxwell fields\cite{bekenstein}, and with an additional external scalar
field $\psi $ minimally coupled to gravity, and universally coupled to
matter, with action given by (SI units are used) 
\begin{eqnarray}
S &=&-\frac{c^{3}}{16\pi G_{0}}\int \sqrt{-g}\phi Rd\Omega +\frac{c^{3}}{%
16\pi G_{0}}\int \sqrt{-g}\frac{\omega \left( \phi \right) }{\phi }\nabla
^{\nu }\phi \nabla _{\nu }\phi d\Omega  \nonumber \\
&&+\frac{c^{3}}{16\pi G_{0}}\int \sqrt{-g}\phi \left[ \frac{1}{2}\nabla
^{\nu }\psi \nabla _{\nu }\psi -U\left( \psi \right) -J\psi \right] d\Omega 
\nonumber \\
&&-\frac{\varepsilon _{0}c}{4}\int \sqrt{-g}\lambda \left( \phi \right)
F_{\mu \nu }F^{\mu \nu }d\Omega -\frac{1}{c}\int \sqrt{-g}j^{\nu }A_{\nu
}d\Omega  \nonumber \\
&&+\frac{1}{c}\int \mathcal{L}_{mat}\left[ \exp \left( \beta \psi \right)
g_{\mu \nu }\right] d\Omega .  \label{SKK}
\end{eqnarray}
In order to have a non-dimensional scalar field $\phi $ of values around
unity, in expression (\ref{SKK}) the constant $G_{0}$ representing Newton
gravitational constant is included, $c$ is the velocity of light in vacuum,
and $\varepsilon _{0}$ is the vacuum permittivity. $\mathcal{L}_{mat}$ is
the lagrangian density of matter, which is assumed to couple to the scalar $%
\psi $. The other symbols are also conventional, $R$ is the Ricci scalar,
and $g$ the determinant of the metric tensor $g_{\mu \nu }$. The Brans-Dicke
parameters $\omega \left( \phi \right) $ is considered a function of $\phi $%
, as it usually results so in the reduction to four dimensions of
multidimensional theories\cite{chavineau}. The function $\lambda \left( \phi
\right) $ in the term of the action of the electromagnetic field is of the
type appearing in Bekenstein's theory and other effective theories\cite%
{mbelek2003}, it does not intervene in the weak field approximation
ultimately employed, but is included for completeness. The electromagnetic
tensor is $F_{\mu \nu }=\nabla _{\mu }A_{\nu }-\nabla _{\nu }A_{\mu }$, \
given in terms of the electromagnetic quadri-vector $A_{\nu }$, with sources
given by the quadri-current $j^{\nu }$. $U$ and $J$ are, respectively, the
potential and source of the field $\psi $. The source $J$ contains
contributions from the matter, electromagnetic field and the scalar $\phi $.
In order to build upon a concrete model we follow the proposal in\cite{MLR}
(convenient dimensional factors differing from those in\cite{MLR} are
employed here)%
\begin{equation}
J=\beta _{mat}\left( \psi ,\phi \right) \frac{8\pi G_{0}}{c^{4}}%
T^{mat}+\beta _{EM}\left( \psi ,\phi \right) \frac{4\pi G_{0}\varepsilon _{0}%
}{c^{2}}F_{\mu \nu }F^{\mu \nu }+\beta _{\phi }\left( \psi ,\phi \right)
T^{\phi },  \label{source}
\end{equation}
where $T^{mat}$ is the trace of the energy-momentum tensor of matter, (note
that this tensor is defined with respect to $g_{\mu \nu }$, not $\exp \left(
\beta \psi \right) g_{\mu \nu }$),\ 
\[
T_{\mu \nu }^{mat}=-\frac{2}{\sqrt{-g}}\frac{\delta \mathcal{L}_{mat}}{%
\delta g^{\mu \nu }}, 
\]%
and $T^{\phi }$ is the trace of the tensor 
\[
T_{\mu \nu }^{\phi }=\nabla _{\mu }\nabla _{\nu }\phi -\nabla ^{\gamma
}\nabla _{\gamma }\phi g_{\mu \nu }+\frac{\omega \left( \phi \right) }{\phi }%
\left( \nabla _{\mu }\phi \nabla _{\nu }\phi -\frac{1}{2}\nabla ^{\gamma
}\phi \nabla _{\gamma }\phi g_{\mu \nu }\right) . 
\]

Variation of (\ref{SKK}) with respect to $g^{\mu \nu }$ results in ($T_{\mu
\nu }^{EM}$ is the usual electromagnetic energy tensor)%
\begin{eqnarray}
\phi \left( R_{\mu \nu }-\frac{1}{2}Rg_{\mu \nu }\right) &=&\frac{8\pi G_{0}%
}{c^{4}}\left[ \lambda \left( \phi \right) T_{\mu \nu }^{EM}+T_{\mu \nu
}^{mat}\right] +T_{\mu \nu }^{\phi }  \nonumber \\
&&+\frac{\phi }{2}\left( \nabla _{\mu }\psi \nabla _{\nu }\psi -\frac{1}{2}%
\nabla ^{\gamma }\psi \nabla _{\gamma }\psi g_{\mu \nu }\right)  \nonumber \\
&&+\frac{\phi }{2}\left( U+J\psi \right) g_{\mu \nu }.  \label{Glm}
\end{eqnarray}

Variation with respect to $\phi $ gives 
\begin{eqnarray*}
\phi R+2\omega \nabla ^{\nu }\nabla _{\nu }\phi &=&\left( \frac{\omega }{%
\phi }-\frac{d\omega }{d\phi }\right) \nabla ^{\nu }\phi \nabla _{\nu }\phi -%
\frac{4\pi G_{0}\varepsilon _{0}}{c^{2}}\phi \frac{d\lambda }{d\phi }F_{\mu
\nu }F^{\mu \nu } \\
&&-\frac{\partial J}{\partial \phi }\psi \phi +\phi \left[ \frac{1}{2}\nabla
^{\nu }\psi \nabla _{\nu }\psi -U\left( \psi \right) -J\psi \right] ,
\end{eqnarray*}
which can be rewritten, using the contraction of (\ref{Glm}) with $g^{\mu
\nu }$ to replace $R$, as 
\begin{eqnarray}
\left( 2\omega +3\right) \nabla ^{\nu }\nabla _{\nu }\phi &=&-\frac{d\omega 
}{d\phi }\nabla ^{\nu }\phi \nabla _{\nu }\phi -\frac{4\pi G_{0}\varepsilon
_{0}}{c^{2}}\phi \frac{d\lambda }{d\phi }F_{\mu \nu }F^{\mu \nu }+\frac{8\pi
G_{0}}{c^{4}}T^{mat}  \nonumber \\
&&+\phi \left[ \frac{1}{2}\nabla ^{\nu }\psi \nabla _{\nu }\psi -U\left(
\psi \right) -J\psi \right] -\frac{\partial J}{\partial \phi }\psi \phi ,
\label{phi}
\end{eqnarray}
where it was used that $T^{EM}=T_{\mu \nu }^{EM}g^{\mu \nu }=0$.

The non-homogeneous Maxwell equations are obtained by varying (\ref{SKK})
with respect to $A_{\nu }$, 
\begin{equation}
\nabla _{\mu }\left\{ \lambda \left( \phi \right) F^{\mu \nu }\right\} =\mu
_{0}j^{\nu }.  \label{Maxwell}
\end{equation}%
with $\mu _{0}$ the vacuum permeability.

The variation with respect to $\psi $ results in 
\begin{equation}
\nabla ^{\nu }\nabla _{\nu }\psi +\frac{1}{\phi }\nabla ^{\nu }\psi \nabla
_{\nu }\phi =-\frac{\partial U}{\partial \psi }-J-\frac{\partial J}{\partial
\psi }\psi +\frac{\beta }{\phi }\frac{8\pi G_{0}}{c^{4}}T^{mat}.  \label{psi}
\end{equation}

Having included $G_{0}$, it is understood that $\phi $ takes values around
its vacuum expectation value (VEV) $\phi _{0}=1$. The scalar $\psi $ is also
dimensionless and of VEV $\psi _{0}$.

Finally, we consider the motion of neutral test particles, coupled to the
scalar $\psi $ as indicated in (\ref{SKK}), which is then obtained requiring
that 
\[
\delta \int mc\sqrt{\exp \left( \beta \psi \right) g_{\mu \nu }dx^{\mu
}dx^{\nu }}=0, 
\]%
to give%
\begin{equation}
\frac{Du^{\gamma }}{Ds}=\frac{\beta }{2}\left( g^{\gamma \nu }-u^{\gamma
}u^{\nu }\right) \partial _{\nu }\psi .  \label{motion}
\end{equation}

It is important to mention that, in order to derive the previous equations,
we have followed the prescription of not varying the trace of the
energy-momentum tensors nor $F_{\mu \nu }F^{\mu \nu }$ in the source term (%
\ref{source}), but only its coefficients $\beta ^{\prime }s$, as done in\cite%
{MLR}. There is no clear reason for doing so, but on the one hand,
inconsistent equations result if the mentioned variations are included. On
the other hand, the exact source term may not depend explicitly on the
tensors considered, and only after the equations are derived and
substitutions made might it be expressible in terms of the say tensors.

\section{Weak-field approximation}

In the weak field approximation, for values of $g_{\mu \nu }$ around $\eta
_{\mu \nu }$ taken as those of flat Minkowski space with signature
(1,-1,-1,-1), so that $g_{\mu \nu }=\eta _{\mu \nu }+h_{\mu \nu }$, we have 
\[
R_{\mu \nu }-\frac{1}{2}R\eta _{\mu \nu }=\frac{1}{2}\left( -\eta ^{\gamma
\delta }\partial _{\gamma \delta }\overline{h}_{\mu \nu }+\partial _{\gamma
\mu }\overline{h}_{\nu }^{\gamma }+\partial _{\gamma \nu }\overline{h}_{\mu
}^{\gamma }-\eta _{\mu \nu }\partial _{\gamma \delta }\overline{h}^{\gamma
\delta }\right) , 
\]
with 
\[
\overline{h}_{\mu \nu }\equiv h_{\mu \nu }-\frac{1}{2}h\eta _{\mu \nu }, 
\]
where 
\[
h\equiv \eta ^{\gamma \delta }h_{\gamma \delta }=-\eta ^{\gamma \delta }%
\overline{h}_{\gamma \delta }. 
\]
The system (\ref{Glm})-(\ref{Maxwell}) can then be written, to lowest order
in $\overline{h}_{\mu \nu }$ and in the perturbations around the VEV's of $%
\phi $ and $\psi $, as 
\begin{equation}
-\eta ^{\gamma \delta }\partial _{\gamma \delta }\overline{h}_{\mu \nu }=%
\frac{16\pi G_{0}}{c^{4}}T_{\mu \nu }^{mat}+2\left( \partial _{\mu \nu }\phi
-\eta ^{\gamma \delta }\partial _{\gamma \delta }\phi \eta _{\mu \nu
}\right) ,  \label{Gik0}
\end{equation}
with the Lorentz gauge 
\begin{equation}
\partial _{\gamma }\overline{h}_{\nu }^{\gamma }=0,  \label{LG}
\end{equation}
\begin{equation}
\left( 2\omega _{0}+3\right) \eta ^{\gamma \delta }\partial _{\gamma \delta
}\phi =\frac{8\pi G_{0}}{c^{4}}T^{mat}-\left. \frac{\partial J}{\partial
\phi }\right\vert _{\phi _{0},\psi _{0}}\psi _{0},  \label{dphi0}
\end{equation}
\begin{equation}
\partial _{\mu }F^{\mu \nu }=\mu _{0}j^{\nu },  \label{Max0}
\end{equation}
\begin{equation}
\eta ^{\gamma \delta }\partial _{\gamma \delta }\psi =\beta \frac{8\pi G_{0}%
}{c^{4}}T^{mat}-\left. \frac{\partial J}{\partial \psi }\right\vert _{\phi
_{0},\psi _{0}}\psi _{0},  \label{dpsi0}
\end{equation}
where $\omega _{0}=\omega \left( \phi _{0}\right) $. It was used in these
equations that, in order to recover the usual physics when the scalar fields
are not excited, one must have $\lambda \left( \phi _{0}\right) =1$, $%
U\left( \psi _{0}\right) =J\left( \psi _{0},\phi _{0}\right) =0$. Also, as
according to\cite{MLR} the contribution form the energy-momentum tensor of
the electromagnetic field through the source $J$ is much larger than all its
other contributions, the latter were neglected in the above equations.

For slow moving neutral masses, the equation (\ref{motion}) corresponds to
the action of a specific force (per unit mass) (Latin indices correspond to
the spatial coordinates) 
\begin{equation}
f_{i}=-\frac{c^{2}}{4}\frac{\partial }{\partial x_{i}}\left( \overline{h}%
_{00}+\overline{h}_{kk}+2\beta \psi \right) +c\frac{\partial \overline{h}%
_{0i}}{\partial t}.  \label{forcepermass}
\end{equation}

Introducing the D\'{}Alembertian operator 
\[
\square =\eta ^{\gamma \delta }\partial _{\gamma \delta }=\frac{1}{c^{2}}%
\frac{\partial ^{2}}{\partial t^{2}}-\nabla ^{2},
\]%
applying it to the force equation (\ref{forcepermass}), and using Eq. (\ref%
{Gik0}), one easily obtains%
\begin{eqnarray*}
\square f_{i} &=&\frac{4\pi G_{0}}{c^{2}}\frac{\partial }{\partial x_{i}}%
\left( T_{00}^{mat}+T_{kk}^{mat}\right) -\frac{16\pi G_{0}}{c^{3}}\frac{%
\partial T_{0i}^{mat}}{\partial t} \\
&&+\frac{c^{2}}{2}\frac{\partial }{\partial x_{i}}\left( \square \phi -\beta
\square \psi -\frac{2}{c^{2}}\frac{\partial ^{2}\phi }{\partial t^{2}}%
\right) .
\end{eqnarray*}%
From Eqs. (\ref{dphi0}) and (\ref{dpsi0}), and retaining only the most
important component $T_{00}^{mat}$, this expression can be conveniently
recast as 
\[
f_{i}=-\frac{\partial \chi }{\partial x_{i}},
\]%
where the "gravitational potential" $\chi $ satisfies%
\begin{eqnarray}
\square \chi  &=&-\frac{4\pi G_{0}}{c^{2}}T_{00}^{mat}+\frac{4\pi G_{0}}{%
c^{2}}\left( \beta ^{2}-\frac{1}{2\omega _{0}+3}\right) T_{00}^{mat} 
\nonumber \\
&&+\frac{\partial ^{2}\phi }{\partial t^{2}}+\frac{c^{2}\psi _{0}}{2}\left( 
\frac{1}{2\omega _{0}+3}\frac{\partial J}{\partial \phi }-\beta \frac{%
\partial J}{\partial \psi }\right) _{\phi _{0},\psi _{0}}.  \label{potential}
\end{eqnarray}%
Expliciting the matter contribution to $\chi $, using expression (\ref%
{source}), one has%
\begin{eqnarray*}
\square \chi  &=&-\frac{4\pi G_{0}}{c^{2}}T_{00}^{mat} \\
&&+\frac{4\pi G_{0}}{c^{2}}\left[ \beta \left( \beta -\psi _{0}\frac{%
\partial \beta _{mat}}{\partial \psi }\right) +\frac{1}{2\omega _{0}+3}%
\left( \psi _{0}\frac{\partial \beta _{mat}}{\partial \phi }-1\right) \right]
_{\phi _{0},\psi _{0}}T_{00}^{mat}+...,
\end{eqnarray*}%
where the dots represent non-matter terms. The first term corresponds to
Newton gravity, while the second term, if one takes $\beta =0$, corresponds
to the matter contribution through the scalar $\phi $, which is constrained
by Solar System tests, requiring large values of $\omega _{0}$. An
interesting conclusion (not to be explored further here) is that the
inclusion of the external scalar $\psi $ could thus allow $\omega _{0}\sim 1$
if $\beta $ is small enough (or, alternatively, if $\beta \simeq \psi
_{0}\partial \beta _{mat}/\partial \psi $), and 
\begin{equation}
\psi _{0}\left. \frac{\partial \beta _{mat}}{\partial \phi }\right\vert
_{\phi _{0},\psi _{0}}\simeq 1.  \label{b2}
\end{equation}%
Note that the condition (\ref{b2}) does not invalidate the conclusions in 
\cite{mbelek2004} as only the term dependent on the matter velocity of the
"force" in the right hand side of (\ref{motion}) is used therein to explain
the dynamics of rotating spiral galaxies.

Making explicit the equation of the scalar $\phi $, Eq. (\ref{dphi0}), with
the expression of the source $J$, Eq. (\ref{source}), one has%
\begin{eqnarray}
\square \phi &=&\frac{8\pi G_{0}}{\left( 2\omega _{0}+3\right) c^{4}}\left(
1-\psi _{0}\left. \frac{\partial \beta _{mat}}{\partial \phi }\right\vert
_{\phi _{0},\psi _{0}}\right) T^{mat}  \nonumber \\
&&-\frac{8\pi G_{0}\varepsilon _{0}}{\left( 2\omega _{0}+3\right) c^{2}}\psi
_{0}\left. \frac{\partial \beta _{EM}}{\partial \phi }\right\vert _{\phi
_{0},\psi _{0}}\left( B^{2}-E^{2}/c^{2}\right) ,  \label{lapphi}
\end{eqnarray}%
where it was used that, in terms of the modulus of the electric and magnetic
vector fields, $E$ and $B$, respectively, one has 
\[
F_{\mu \nu }F^{\mu \nu }=2\left( B^{2}-E^{2}/c^{2}\right) , 
\]%
and where the contribution from $\phi $ itself as its source was not
considered because, even if it is present, in the weak-field approximation
one has 
\[
T^{\phi }=-3\square \phi , 
\]%
and so its effect amounts to a redefinition of the rest of the coefficients
in the equations for $\phi $ and $\psi $.

A point worth noting is that condition (\ref{b2}) refers so far to the
motion of massive bodies, while the most stringent bounds on $\omega _{0}$
come from the propagation of electromagnetic waves near the Sun\cite%
{bertotti}, not affected by the coupling of $\psi $ to matter. According to
the expression (\ref{lapphi}) these bounds can also be accommodated, always
with $\omega _{0}\sim 1$, if the same condition (\ref{b2}) holds.

With all this, the contributions other than the matter to the potential $%
\chi $ can then be obtained from (\ref{potential}) as (we write $\chi =\chi
_{mat}+\chi ^{\prime }$) 
\begin{equation}
\square \chi ^{\prime }=\frac{\partial ^{2}\phi }{\partial t^{2}}+4\pi
G_{0}\varepsilon _{0}\psi _{0}\left( \frac{1}{2\omega _{0}+3}\frac{\partial
\beta _{EM}}{\partial \phi }-\beta \frac{\partial \beta _{EM}}{\partial \psi 
}\right) _{\phi _{0},\psi _{0}}\left( B^{2}-E^{2}/c^{2}\right) .
\label{potem}
\end{equation}

In\cite{MLR} it is argued that in order to explain discordant measurements
of $G=G_{0}/\phi $ as due to the $\phi $ generated by the Earth's magnetic
field according to (\ref{lapphi}), one must have%
\begin{equation}
\frac{8\pi G_{0}\varepsilon _{0}}{\left( 2\omega _{0}+3\right) c^{2}}\psi
_{0}\left. \frac{\partial \beta _{EM}}{\partial \phi }\right\vert _{\phi
_{0},\psi _{0}}=-\left( 5.4\pm 0.6\right) \times 10^{-8}\frac{A^{2}}{N^{2}},
\label{mbelek}
\end{equation}%
while the value of $\beta \left. \partial J/\partial \psi \right\vert _{\phi
_{0},\psi _{0}}$ does not enter the equation of $\phi $ and is thus left
unspecified. In the following we will evaluate the force predicted by (\ref%
{potem}) for the resonant electromagnetic field in a conical cavity,
assuming that the coefficient in the brackets in (\ref{potem}) can be
estimated from the value (\ref{mbelek}) alone.

\section{Normal modes in a conical cavity}

As done in\cite{egan} we consider a conical cavity with side walls
corresponding to a truncated cone, with spherical sections as end caps. The
cone axis is taken as the $z$ direction, the lateral wall corresponds to the
spherical angle $\theta =\theta _{0}$ (half angle of the cone), and the
spherical caps to the radii $r=r_{1,2}$, with $r_{2}>r_{1}$.

The resonant modes correspond to standing electromagnetic waves satisfying
the vector wave equation ($\mathbf{F}$ stands for either the electric field $%
\mathbf{E}$, or the magnetic induction $\mathbf{B}$)%
\[
\frac{1}{c^{2}}\frac{\partial ^{2}\mathbf{F}}{\partial t^{2}}-\nabla ^{2}%
\mathbf{F}=0. 
\]%
The modes with rotational symmetry and $\mathbf{B}$ transverse to the $z$
direction $\mathbf{e}_{z}$ (called the TM modes) that satisfy this equation
are (spherical coordinates are employed, with unit vectors $\mathbf{e}_{r}$, 
$\mathbf{e}_{\theta }$\ and $\mathbf{e}_{\varphi }$) (see\cite{egan} and
references therein for details) 
\begin{eqnarray}
\mathbf{B} &=&-CkR\left( r\right) Q^{\prime }\left( \theta \right) \cos
\left( \omega t\right) \mathbf{e}_{\varphi },  \label{bmode} \\
\mathbf{E}/c &=&C\left\{ \frac{R\left( r\right) }{r}n\left( n+1\right)
Q\left( \theta \right) \mathbf{e}_{r}\right.  \nonumber \\
&&\left. +\left[ \frac{R\left( r\right) }{r}+R^{\prime }\left( r\right) %
\right] Q^{\prime }\left( \theta \right) \mathbf{e}_{\theta }\right\} \sin
\left( \omega t\right)  \label{emode}
\end{eqnarray}%
where $C$ is a global constant. The functions $R$ and $Q$ are defined as 
\begin{eqnarray*}
Q\left( \theta \right) &=&P_{n}\left( \cos \theta \right) , \\
R\left( r\right) &=&R_{+}\left( r\right) \cos \alpha +R_{-}\left( r\right)
\sin \alpha , \\
R_{\pm }\left( r\right) &=&\frac{J_{\pm \left( n+1/2\right) }\left(
kr\right) }{\sqrt{r}},
\end{eqnarray*}%
where $P_{n}$ is the Legendre polynomial of order $n$, $J_{m}$ the Bessel
function of the first kind of order $m$, and $\alpha $ and $k$ constants to
be determined along with the order $n$. By construction, the magnetic field
satisfies the boundary condition of zero normal component at the metallic
walls, while in order to have zero tangential components of the electric
field at the walls, the order $n$ of the Legendre polynomial must satisfy 
\[
P_{n}\left( \cos \theta _{0}\right) =0, 
\]%
the wavenumber $k$ the condition%
\[
\left[ \frac{R_{+}}{r}+R_{+}^{\prime }\right] _{r_{2}}\left[ \frac{R_{-}}{r}%
+R_{-}^{\prime }\right] _{r_{1}}=\left[ \frac{R_{+}}{r}+R_{+}^{\prime }%
\right] _{r_{1}}\left[ \frac{R_{-}}{r}+R_{-}^{\prime }\right] _{r_{2}}, 
\]%
and $\alpha $%
\[
\tan \alpha =-\frac{R_{+}\left( r_{2}\right) /r_{2}+R_{+}^{\prime }\left(
r_{2}\right) }{R_{-}\left( r_{2}\right) /r_{2}+R_{-}^{\prime }\left(
r_{2}\right) }. 
\]%
The resonant mode angular frequency is thus determined as $\omega =kc$.

There exists a complementary set of modes with $\mathbf{E}$ transverse to
the $z $ direction (TE modes), but for concreteness we study only the lowest
frequency TM modes.

An important parameter is the quality factor of the cavity, $Q_{cav} $, for
each mode. It is conventionally defined as 
\begin{equation}
Q_{cav}\equiv \frac{\omega \left\langle U\right\rangle }{\left\langle
W\right\rangle },  \label{quw}
\end{equation}%
where $\omega $ is the angular frequency of the mode, $\left\langle
U\right\rangle $ is the temporal average of its electromagnetic energy, and $%
\left\langle W\right\rangle $ is the average dissipated power in the wall
cavities. As the average electric energy is equal to the average magnetic
energy in the cavity, and the loss power can obtained from the value of the
magnetic field at the boundary, an explicit, practical expression of $%
Q_{cav} $ can be obtained in terms of solely the magnetic field as\cite%
{jackson} 
\[
Q_{cav}=\frac{2}{\delta }\frac{\int \left\langle B^{2}\right\rangle dV}{\int
\left\langle B^{2}\right\rangle dS}, 
\]%
where the integrals are extended to the volume and the internal surface of
the cavity, respectively, and $\delta $ is the penetration length in the
metal wall, of resistivity $\eta $,%
\begin{equation}
\delta =\sqrt{\frac{2\eta }{\mu _{0}\omega }}.  \label{pdepth}
\end{equation}%
From (\ref{bmode}) one can thus write%
\begin{equation}
Q_{cav}=\frac{2}{\delta }\frac{\int \left[ R\left( r\right) Q^{\prime
}\left( \theta \right) \right] ^{2}dV}{\int \left[ R\left( r\right)
Q^{\prime }\left( \theta \right) \right] ^{2}dS}.  \label{qfin}
\end{equation}

If the cavity is fed with an average electromagnetic power $P$, in the
permanent regime one has $\left\langle W\right\rangle =P$, and so, from (\ref%
{bmode}) and (\ref{quw}), 
\begin{equation}
\left\langle U\right\rangle =\frac{\int \left\langle B^{2}\right\rangle dV}{%
\mu _{0}}=\frac{C^{2}k^{2}}{2\mu _{0}}\int \left[ R\left( r\right) Q^{\prime
}\left( \theta \right) \right] ^{2}dV=\frac{Q_{cav}P}{\omega },
\label{csquare}
\end{equation}%
which allows to determine the global constant $C$, given the fed average
power and the characteristics of the cavity for the considered mode.

\section{Force on the cavity}

In the permanent regime of the established resonant mode, sustained against
decay by a continuous power input $P$, the electromagnetic field (\ref{bmode}%
)-(\ref{emode}) corresponds to%
\begin{eqnarray}
B^{2}-E^{2}/c^{2} &=&F_{B}\left( r,\theta \right) \cos ^{2}\left( \omega
t\right) -F_{E}\left( r,\theta \right) \sin ^{2}\left( \omega t\right) 
\nonumber \\
&=&\frac{1}{2}\left( F_{B}-F_{E}\right) +\frac{1}{2}\left(
F_{B}+F_{E}\right) \cos \left( 2\omega t\right) ,  \label{b2me2}
\end{eqnarray}%
where%
\begin{eqnarray}
F_{B}\left( r,\theta \right) &=&C^{2}k^{2}\left[ R\left( r\right) Q^{\prime
}\left( \theta \right) \right] ^{2},  \label{FB} \\
F_{E}\left( r,\theta \right) &=&C ^{2}\left\{\left[ \frac{R\left( r\right) }{%
r}n\left( n+1\right) Q\left( \theta \right) \right] ^{2}\right.  \nonumber \\
&&\left. +\left[ \frac{R\left( r\right) }{r}+R^{\prime }\left( r\right) %
\right] ^{2}Q^{\prime 2}\left( \theta \right) \right\} .  \label{FE}
\end{eqnarray}

From (\ref{lapphi}), expression (\ref{b2me2}) then leads to a constant plus
an harmonic in time contribution to $\phi $, which, together with (\ref%
{b2me2}) in (\ref{potem}), result in $\chi ^{\prime }$ also having a
constant plus an harmonic part. The latter has a zero contribution to the
time average of the force, and so we consider only the constant part, $\chi
_{0}^{\prime }$, whose equation is, from (\ref{potem}),%
\begin{equation}
\nabla ^{2}\chi _{0}^{\prime }=\varkappa \left( F_{B}-F_{E}\right) ,
\label{lapchi}
\end{equation}%
where, using (\ref{mbelek}),%
\begin{eqnarray}
\varkappa &=&-2\pi G_{0}\varepsilon _{0}\psi _{0}\left( \frac{1}{2\omega
_{0}+3}\frac{\partial \beta _{EM}}{\partial \phi }-\beta \frac{\partial
\beta _{EM}}{\partial \psi }\right) _{\phi _{0},\psi _{0}}  \nonumber \\
&\simeq &-\frac{2\pi G_{0}\varepsilon _{0}\psi _{0}}{2\omega _{0}+3}\left. 
\frac{\partial \beta _{EM}}{\partial \phi }\right\vert _{\phi _{0},\psi
_{0}}\simeq 1.2\times 10^{9}\left( \frac{Am}{Ns}\right) ^{2}.  \label{kconst}
\end{eqnarray}

Note that the magnetic field in the right-hand side of (\ref{potem}) is the
total field, which includes the contribution from the Earth's magnetic
field. The latter, although of much smaller magnitude than that of the
cavity cannot be neglected due to its large spatial scale. However, one has
for the time average (denoted by $\left\langle ...\right\rangle $) 
\begin{eqnarray*}
\left\langle B_{Earth}^{2}+B_{cavity}^{2}\right\rangle
&=&B_{Earth}^{2}+\left\langle B_{cavity}^{2}\right\rangle +2\mathbf{B}%
_{Earth}\cdot \left\langle \mathbf{B}_{cavity}\right\rangle \\
&=&B_{Earth}^{2}+F_{B}/2,
\end{eqnarray*}%
since $\left\langle \mathbf{B}_{cavity}\right\rangle =0$. In this way, the
contribution from the magnetic fields of the Earth and of the cavity to the
potential $\chi _{0}^{\prime }$ can be separated, and that of the cavity
alone is correctly described by (\ref{lapchi}).

Eq. (\ref{lapchi}) is solved taking into account that its right-hand side is
zero outside the cavity, so that, using the axial symmetry, the solution of
Poisson equation (\ref{lapchi}) is 
\begin{eqnarray}
\chi _{0}^{\prime }\left( r,\theta \right) &=&-\frac{\varkappa }{\pi }\int 
\frac{F_{B}\left( r^{\prime },\theta ^{\prime }\right) -F_{E}\left(
r^{\prime },\theta ^{\prime }\right) }{\sqrt{r^{2}+r^{\prime 2}-2rr^{\prime
}\cos \left( \theta ^{\prime }-\theta \right) }}  \nonumber \\
&&\times K\left( -\frac{4rr^{\prime }\sin \theta \sin \theta ^{\prime }}{%
r^{2}+r^{\prime 2}-2rr^{\prime }\cos \left( \theta ^{\prime }-\theta \right) 
}\right) r^{\prime 2}\sin \theta ^{\prime }dr^{\prime }d\theta ^{\prime }
\label{chi0sol}
\end{eqnarray}
where $K$ is the complete elliptic integral of the first kind, and the
integral is extended to the interior of the cavity. Note that as the volume
integral of the left-hand side of (\ref{lapchi}) is equal to zero, $\mathbf{%
\nabla }\chi _{0}^{\prime }$ decays rapidly outside the cavity.

Assuming a cavity with thin walls (but much thicker than the penetration
depth $\delta $, in order to the boundary conditions used to be correct) of
mass surface density $\sigma $, the force on the cavity is finally evaluated
as%
\begin{equation}
\mathbf{F}=-\sigma \int \mathbf{\nabla }\chi _{0}^{\prime }dS,
\label{forcecavity}
\end{equation}%
where the integral is extended to the internal surface of the cavity. Due to
the axial symmetry the force has only a $z$ component%
\begin{eqnarray*}
F_{z} &=&-\sigma \int \frac{\partial \chi _{0}^{\prime }}{\partial z}%
dS=-\sigma \int \mathbf{\nabla }\chi _{0}^{\prime }\cdot \mathbf{e}_{z}dS, \\
&=&-\sigma \int \left( \cos \theta \frac{\partial \chi _{0}^{\prime }}{%
\partial r}-\frac{\sin \theta }{r}\frac{\partial \chi _{0}^{\prime }}{%
\partial \theta }\right) dS,
\end{eqnarray*}%
which is explicitly written as%
\begin{eqnarray}
-\frac{F_{z}}{2\pi \sigma } &=&r_{2}^{2}\int_{0}^{\theta _{0}}\left. \left(
\cos \theta \frac{\partial \chi _{0}^{\prime }}{\partial r}-\frac{\sin
\theta }{r}\frac{\partial \chi _{0}^{\prime }}{\partial \theta }\right)
\right\vert _{r_{2}}\sin \theta d\theta  \nonumber \\
&&+r_{1}^{2}\int_{0}^{\theta _{0}}\left. \left( \cos \theta \frac{\partial
\chi _{0}^{\prime }}{\partial r}-\frac{\sin \theta }{r}\frac{\partial \chi
_{0}^{\prime }}{\partial \theta }\right) \right\vert _{r_{1}}\sin \theta
d\theta  \nonumber \\
&&+\sin \theta _{0}\int_{r_{1}}^{r_{2}}\left. \left( \cos \theta \frac{%
\partial \chi _{0}^{\prime }}{\partial r}-\frac{\sin \theta }{r}\frac{%
\partial \chi _{0}^{\prime }}{\partial \theta }\right) \right\vert _{\theta
_{0}}rdr.  \label{fzfin}
\end{eqnarray}

There are no details in the literature as to the precise dimensions of the
cavities used in the experiments, so that an example roughly similar to the
overall dimension reported and with the proportions observed in the
published photographs will be used. Assuming a wall of thickness 1 mm, and a
copper mass density of $8.9\times 10^{3}\,kg/m^{3}$, we have $\sigma
=8.9\,kg/m^{2}$.

We further consider the copper cavity to have $r_{1}=18$ cm, $r_{2}=36$ cm,
and $\theta _{0}=22%
{{}^\circ}%
$. For this cavity, the lowest TM mode corresponds to the order $n=5.75632$
of the Legendre polynomial, with a resonant frequency $\nu =1.05\,GHz$. For
a resistivity $\eta =1.72\times 10^{-8}\,\Omega \,m$ the quality factor for
this mode is $Q_{cav}=3.13\times 10^{4}$. The next two TM modes have the
same order $n=5.75632$, and resonant frequencies $\nu =2.05\,GHz$ and $\nu
=2.76\,GHz$, with quality factors $Q_{cav}=3.11\times 10^{4}$ and $%
Q_{cav}=5.24\times 10^{4}$, respectively.

For an average power $P=1\,kW$, the constant $C$ is evaluated for each mode
using (\ref{csquare}), and (\ref{FB}) and (\ref{FE}) used in (\ref{chi0sol})
to obtain by numerical integration the values of $\chi _{0}^{\prime }\left(
r,\theta \right) $ needed in the numerical evaluation of (\ref{fzfin}).

Note that, from (\ref{csquare}), the force on the cavity is proportional to
the fed power, and to the quality factor $Q_{cav}$.

For the lowest TM mode ($\nu =1.05\,GHz$) the value obtained is $%
F_{z}=7.7\,N $, while for the next two TM modes, with $\nu =2.05\,GHz$ and $%
\nu =2.76\,GHz $, we obtained $F_{z}=-1.4\,N$ and $F_{z}=-0.9N$,
respectively. The values reported in\cite{juan} are not easy to compare with
as the power of the microwave source is distributed over a rather wide range
of frequencies, so that the actual power into the resonant mode is not
precisely defined. Using a spectrum analysis of the power source the authors
evaluate, for instance, that when $F_{z}=-0.3\,N$ the actual power into the
resonant mode is $P=0.12\,kW$,\ \ which would correspond to $F_{z}=-2.5\,N$
at $P=1\,kW$. The last two modes considered are closer to the reported value
of the resonance, $\nu =2.45\,GHz$, and give theoretical results with the
correct sign and similar magnitude. As according to the model the force is
proportional to the thickness of the wall, depends also on the precise
geometry of the cavity (neither of them reported in the literature), and as
the value (\ref{kconst}) is only an estimation, since the contribution from $%
\beta \left. \partial \beta _{EM}/\partial \psi \right\vert _{\phi _{0},\psi
_{0}}$ cannot be ascertained independently, the results seem consistent with
the measured force being due to the studied effect.

Note that the lowest mode ($\nu =1.05\,GHz$) leads to a force much larger in
magnitude and of opposite direction to that of the next two modes. This and
other dependencies of the predicted force, as the proportionality to the
cavity wall thickness (within certain limits as $\mathbf{\nabla }\chi
_{0}^{\prime }$ decays rapidly outside the cavity), can be explored
experimentally with relative ease to test the theory.

Finally, it is worth noting that the weak energy condition (WEC)\cite%
{hawking} is violated for the cavity, as is the case in other models of
propellant-less drive\cite{lobo}. In effect, from (\ref{Gik0}), the WEC is
written for the cavity 
\begin{equation}
\left( \partial _{\mu \nu }\phi -\eta ^{\gamma \delta }\partial _{\gamma
\delta }\phi \eta _{\mu \nu }\right) U^{\mu }U^{\nu }\geq 0,  \label{wec}
\end{equation}%
for any time-like four-vector $U^{\mu }$. By taking $U^{\mu }=\left(
1,0,0,0\right) $ one has the particular WEC 
\[
\nabla ^{2}\phi \geq 0, 
\]
which is seen from (\ref{lapphi}) and (\ref{b2me2}) to be violated at
different times and regions inside the cavity.

\section{Discussion}

It was shown that the weak field approximation of a rather general
scalar-tensor theory of gravity, which includes an additional scalar with
strong coupling to the electromagnetic field, as proposed in\cite{MLR},
could account for the forces reported on asymmetric resonant cavities.
Although highly speculative, it is interesting that this was done using the
same coupling coefficient adjusted by\cite{MLR} to explain discordant
measurements of Newton gravitational constant. It is also of interest that
the inclusion of the external scalar $\psi $ can help to reconcile the Solar
System tests with values of the Brans-Dicke parameter $\omega $ close to
unity (see relation (\ref{b2})). The weakest part of the theory seems to be
that there is no clear way of preventing large gravitational effects due to
the magnetic field of the Earth, as predicted by Eq. (\ref{potem}). A
possible solution can be sought in non-linear effects, such as those due to
the second terms in the left-hand sides of (\ref{phi}) and (\ref{psi}). In
effect, their inclusion would modify (\ref{potential}) to%
\[
\square \chi =\left( \square \chi \right) _{original}+\frac{c^{2}}{2}\left(
\omega _{0}^{\prime }\partial ^{\nu }\phi \partial _{\nu }\phi -\frac{1}{2}%
\partial ^{\nu }\psi \partial _{\nu }\psi -\beta \partial ^{\nu }\phi
\partial _{\nu }\psi \right) , 
\]%
where $\omega _{0}^{\prime }\equiv \left( d\omega /d\phi \right) _{\phi
_{0},\psi _{0}}$. For the stationary case of the Earth's magnetic field one
would then have%
\begin{eqnarray}
\nabla ^{2}\chi &=&\frac{4\pi G_{0}}{c^{2}}T_{00}^{mat}-\frac{4\pi G_{0}}{%
c^{2}}\left( \beta ^{2}-\frac{1}{2\omega _{0}+3}\right) T_{00}^{mat} 
\nonumber \\
&&-\frac{c^{2}\psi _{0}}{2}\left( \frac{1}{2\omega _{0}+3}\frac{\partial J}{%
\partial \phi }-\beta \frac{\partial J}{\partial \psi }\right) _{\phi
_{0},\psi _{0}}  \nonumber \\
&&+\frac{c^{2}}{2}\left( \omega _{0}^{\prime }\mathbf{\nabla }\phi \cdot 
\mathbf{\nabla }\phi -\frac{1}{2}\mathbf{\nabla }\psi \cdot \mathbf{\nabla }%
\psi -\beta \mathbf{\nabla }\phi \cdot \mathbf{\nabla }\psi \right) .
\label{chi_nonlineal}
\end{eqnarray}%
If the terms $\mathbf{\nabla }\phi \cdot \mathbf{\nabla }\phi $, $\mathbf{%
\nabla }\psi \cdot \mathbf{\nabla }\psi $ and $\mathbf{\nabla }\phi \cdot 
\mathbf{\nabla }\psi $ were to dominate over $\nabla ^{2}\phi $ and $\nabla
^{2}\psi $, respectively, the equations (\ref{phi}) and (\ref{psi}) would
result in%
\begin{eqnarray*}
\omega _{0}^{\prime }\mathbf{\nabla }\phi \cdot \mathbf{\nabla }\phi -\frac{1%
}{2}\mathbf{\nabla }\psi \cdot \mathbf{\nabla }\psi &\simeq &-\frac{8\pi
G_{0}}{c^{4}}T^{mat}+\left. \frac{\partial J}{\partial \phi }\right\vert
_{\phi _{0},\psi _{0}}\psi _{0}, \\
\mathbf{\nabla }\phi \cdot \mathbf{\nabla }\psi &\simeq &-\beta \frac{8\pi
G_{0}}{c^{4}}T^{mat}+\left. \frac{\partial J}{\partial \psi }\right\vert
_{\phi _{0},\psi _{0}}\psi _{0},
\end{eqnarray*}%
which clearly cancel the terms in (\ref{chi_nonlineal}) leading to large
values of the force. Put more plainly, $\nabla ^{2}\phi $ and $\nabla
^{2}\psi $ are sources of the potential $\chi ^{\prime }$ and so situations
where they are small or even zero would reduce the gravitational effect of
the electromagnetic field. Note that in the case of a static magnetic field
outside its sources one can write $\mathbf{B}=\mathbf{\nabla }\Psi $, with $%
\nabla ^{2}\Psi =0$, so it is possible that equations like (\ref{phi}) and (%
\ref{psi}) for the static case%
\begin{eqnarray*}
\left( 2\omega _{0}+3\right) \nabla ^{2}\phi +\omega _{0}^{\prime }\mathbf{%
\nabla }\phi \cdot \mathbf{\nabla }\phi -\frac{1}{2}\mathbf{\nabla }\psi
\cdot \mathbf{\nabla }\psi &\propto &B^{2}=\mathbf{\nabla }\Psi \cdot 
\mathbf{\nabla }\Psi , \\
\nabla ^{2}\psi +\mathbf{\nabla }\phi \cdot \mathbf{\nabla }\psi &\propto
&B^{2}=\mathbf{\nabla }\Psi \cdot \mathbf{\nabla }\Psi ,
\end{eqnarray*}%
have the solutions $\mathbf{\nabla }\phi \propto \mathbf{\nabla }\psi
\propto \mathbf{\nabla }\Psi $, and $\nabla ^{2}\phi =\nabla ^{2}\psi =0$,
which is in general not possible for the case of the cavity, for which the
constant part of $B^{2}-E^{2}/c^{2}$ cannot be written as $\mathbf{\nabla }%
\Psi \cdot \mathbf{\nabla }\Psi $.

If this is what happens in the case of the Earth's magnetic field, it would
seem to invalidate the derivations in\cite{MLR}, where the solution with $%
\nabla ^{2}\phi \neq 0$ was used. However, it can be shown that if $\omega
_{0}^{\prime }/\left( 2\omega _{0}+3\right) \sim 1$, both solutions are
numerically similar.

Note that if this is possible for a static magnetic field, it could possibly
not be the case for a static electric field outside its sources, which also
satisfies $\mathbf{E}=\mathbf{\nabla }\Psi $, with $\nabla ^{2}\Psi =0$,
because the difference of sign would not allow real solutions. However, it
can be readily shown that also the electrostatic case have solutions of the
type considered if $\omega _{0}^{\prime }>0$. It is so expected that static
magnetic and electric fields show no unusual gravitational effects, while
non-stationary electromagnetic fields do. Along these lines note finally
that $\partial ^{2}\phi /\partial t^{2}$is also a source of the potential $%
\chi ^{\prime }$, which would contribute in transient situations.

\section*{Appendix: Modified Maxwell equations}

The generalized Maxwell equations, Eq. (\ref{Maxwell}), are 
\[
\nabla _{\nu }\left[ \lambda \left( \phi \right) F^{\mu \nu }\right] =-\mu
_{0}j^{\mu }. 
\]
Since 
\begin{eqnarray*}
\nabla _{\nu }\left[ \lambda \left( \phi \right) F^{\mu \nu }\right] &=&%
\frac{1}{\sqrt{-g}}\frac{\partial }{\partial x^{\nu }}\left[ \sqrt{-g}%
\lambda \left( \phi \right) F^{\mu \nu }\right] \\
&=&\lambda \left( \phi \right) \left[ \frac{\partial F^{\mu \nu }}{\partial
x^{\nu }}+F^{\mu \nu }\frac{\partial }{\partial x^{\nu }}\ln \left( \lambda 
\sqrt{-g}\right) \right] ,
\end{eqnarray*}%
one has 
\[
\frac{\partial F^{\mu \nu }}{\partial x^{\nu }}=-\frac{\mu _{0}}{\lambda
\left( \phi \right) }j^{\mu }-F^{\mu \nu }\frac{\partial }{\partial x^{\nu }}%
\ln \left( \lambda \sqrt{-g}\right) . 
\]

Keeping up to first order in the excitations of the scalars one has 
\[
\frac{\partial F^{\mu \nu }}{\partial x^{\nu }}=-\mu _{0}\left[ 1-\lambda
_{0}^{\prime }\left( \phi -\phi _{0}\right) \right] j^{\mu }-F^{\mu \nu } 
\frac{\partial }{\partial x^{\nu }}\left( \lambda _{0}^{\prime }\phi - 
\overline{h}/2\right) , 
\]
with 
\[
\lambda _{0}^{\prime }\equiv \left. \frac{d\lambda }{d\phi }\right\vert
_{\phi _{0}}. 
\]

The equations for $\phi $ and $\overline{h}$ are at the same order of
approximation (with only electromagnetic sources) 
\begin{eqnarray*}
\square \phi &=&-\frac{8\pi G_{0}\varepsilon _{0}}{\left( 2\omega
_{0}+3\right) c^{2}}\psi _{0}\left. \frac{\partial \beta _{EM}}{\partial
\phi }\right\vert _{\phi _{0},\psi _{0}}\left( B^{2}-E^{2}/c^{2}\right) , \\
\square \overline{h} &=&\eta ^{\mu \nu }\square \overline{h}_{\mu \nu
}=6\square \phi ,
\end{eqnarray*}
so that, if one defines $\Theta \equiv \lambda _{0}^{\prime }\phi -\overline{%
h}/2$, its equation is 
\begin{equation}
\square \Theta =-\frac{8\pi G_{0}\varepsilon _{0}\left( \lambda _{0}^{\prime
}-3\right) }{\left( 2\omega _{0}+3\right) c^{2}}\psi _{0}\left. \frac{%
\partial \beta _{EM}}{\partial \phi }\right\vert _{\phi _{0},\psi
_{0}}\left( B^{2}-E^{2}/c^{2}\right) ,  \label{theta}
\end{equation}%
while the Maxwell equations can be written in usual vector form as
(neglecting the order one correction $\lambda _{0}^{\prime }\left( \phi
-\phi _{0}\right) $ to the four-current $j^{\mu }$) 
\begin{eqnarray*}
\mathbf{\nabla }\cdot \mathbf{E} &=&\frac{\rho }{\varepsilon _{0}}-\mathbf{%
\nabla }\Theta \cdot \mathbf{E}, \\
\mathbf{\nabla }\times \mathbf{E} &=&-\frac{\partial \mathbf{B}}{\partial t}%
,\;\;\;\mathbf{\nabla }\cdot \mathbf{B}=0, \\
\mathbf{\nabla }\times \mathbf{B} &=&\mu _{0}\mathbf{j}+\frac{1}{c^{2}}\frac{%
\partial \mathbf{E}}{\partial t}+\frac{1}{c^{2}}\frac{\partial \Theta }{%
\partial t}\mathbf{E}-\mathbf{\nabla \Theta }\times \mathbf{B}.
\end{eqnarray*}
These equations, together with Eq. (\ref{theta}), form a complete set of for
the electromagnetic field.

It is interesting that in\cite{MLR} the function $\lambda \left( \phi
\right) $ obtained from reduction to four dimensions of the particular
Kaluza-Klein theory employed is $\lambda \left( \phi \right) =\phi ^{3}$, so
that $\lambda _{0}^{\prime }=3$, which according to Eq. (\ref{theta})
results in $\Theta =0$ (the value of $\Theta $ is assumed zero at spatial
infinity). Up to first order the Maxwell equations are so the usual ones in
this case. For different theories, however, $\lambda _{0}^{\prime }\neq 3$,
and the Maxwell equations can be modified. Note that even the case with no
direct coupling, $\lambda \left( \phi \right) =1$, leads to modified Maxwell
equations, which comes directly form the modification of the metric
(relative to Minkowski space) due to the scalar field.


\begin{thebibliography}{99}
\bibitem{juan} Yang J., Wang Y.-Q., Li P.-F., Wang Y., Wang Y.-M., Ma Y.-J.
2012, \textit{Acta Phys. Sin.} \textbf{61} 110301.

\bibitem{emdrive} http://www.emdrive.com

\bibitem{lobo} Lobo F. S. N. and Visser M. 2004, \textit{Class. Quantum Grav.%
} \textbf{21} 5871.

\bibitem{barcelo} Barcelo C and Visser M. 2000, \textit{Class. Quantum Grav.}
\textbf{17} 3843.

\bibitem{MLR} Mbelek J. P. and Lachi\`{e}ze-Rey M. 2002, \textit{Grav. \&
Cosmol.} \textbf{8} 331.

\bibitem{Mb} Mbelek J. P. 2004, \textit{Grav. \& Cosmol.} \textbf{10} 233.

\bibitem{mbelek2003} Mbelek J. P. and Lachi\`{e}ze-Rey M. 2003, \textit{%
Astron. Astrophys.} \textbf{397} 803.

\bibitem{mbelek2004} Mbelek J. P. 2004, \textit{Astron. Astrophys.} \textbf{%
424} 761.

\bibitem{hawking} Hawking S. W. and Ellis G. F. R., \textit{The Large Scale
Structure of Spacetime} (Cambridge University Press, Cambridge, England,
1973).

\bibitem{bransdicke} Brans C., Dicke R. H. 1961 \textit{Phys. Rev.} \textbf{%
124} 925.

\bibitem{bekenstein} Bekenstein J. D. 1982 \textit{Phys. Rev. D} \textbf{25}
1527.

\bibitem{chavineau} Chavineau, B. 2007 \textit{Phys. Rev. D} \textbf{76}
104023.

\bibitem{bertotti} Bertotti B., Iess L., Tortora P. 2003 \textit{Nature} 
\textbf{425} 374.

\bibitem{egan} http://www.gregregan.net/SCIENCE/Cavity/Cavity.html

\bibitem{jackson} Jackson J. D., \textit{Classical Electrodynamics, Third
Edition} (John Wiley \& Sons, New York, 1999)
\end{thebibliography}
\end{document}